# Formation of aeroplane-like ZnO and tetrahexahedral Pt nanocrystals: systemic symmetry and symmetry breaking during nanocrystallization

Yan Zhou[1,2,3], Junyan Zhang[1*], Bin Zhang[1], Jutao Jin[1], Aimin Liang[1], Yuqing Da[3] and Jiangong Li[2]

[1]State Key Laboratory of Solid Lubrication, Lanzhou Institute of Chemical Physics, Chinese Academy of Sciences, Lanzhou 730000, China

[2] Institute of Materials Science and Engineering, Lanzhou University, Lanzhou 730000, China

[3]Tianhua Institute of Chemical Machinery & Automation, Lanzhou 730060, China

*e-mail: junyanzh@yahoo.com

Symmetries are so involved in crystallographic detail that it is indispensable when crystal cell or lattice is mentioned. For nanocrystals, however, many are 'systemic symmetries', meaning that the whole structures of nanocrystal particles are symmetrical from a different view. In that case, the development of the systemic symmetry is dictated by the law of symmetry breaking; that is, nanocrystallization does, to some extent, exceed current theories associated with crystallography.

**1.  Introduction: systemic symmetry and symmetry breaking during crystallization**

The aeroplane-like ZnO nanocrystals and the tetrahexahedral Pt Nanocrystals are cited here as examples of the crystal structures with high symmetries such that the particles are very symmetrical in our immediate visual perception. We refer to this property as 'systemic symmetries'. For crystals, systemic symmetry is neither similar to the symmetries of the cell nor to that of the lattice. More generally, the symmetries of crystals involve 32 point groups to describe the symmetry of the cell, or 230 space groups to describe the space translation symmetry of the lattice. As for the systemic symmetries of crystals, the whole relative isolated structures may contain symmetry elements including inversion centers, rotation axes or symmetry planes, and the symmetry operations about these symmetry elements may transform crystals into states coincidental with the starting state.

In fact, systemic symmetry is perceived commonly in people's life, for it arouses people's esthetic pleasing and original perceive about native crystals that some perfect minerals are always symmetrical with regular geometric shapes, and this fascinating property was summarized partly as constancy of interfacial angles by Nicholas Steno in 1669. About 160 years later, group theory, which should have summarized the rest part of the fascinating properties for crystals, was born. Unfortunately, microscopy diverted people's interesting to the micro structures of crystals during the following years, and the symmetry of cell and lattice had almost attracted all attention. Systemic symmetry of crystals has not been recognized again until nanocrystals occur. As we know, many micro-nanocrystals tend to possess high systemic symmetries with complex shapes, such as nano-polyhedron, nano-flowers, and so on. Thus, scientists ought to consider systemic symmetry as another important section just as the symmetry of cell or lattice.

When a nanocrystal is systemic symmetrical, it could be characterized by a point group, which symbolizes a set of symmetrical operations and symmetrical elements meeting at one point at least. To show the systemic symmetry, a 2-dimensional (2D) Bravais cell with $C_6$ symmetry, which possesses six-fold rotational symmetry just as a regular hexagon, is constructed in Fig. 1. A series of 2D crystals with different systemic symmetrical elements, including symmetry planes, inversion centers, proper axes and improper axes, could be built based on the same cell and lattice (Fig. 1). The amorphous wafer shown in Fig. 1 is considered to possess continuous rotation symmetry for its isotropy, and this is the extremely high symmetry that a 2D structure could possess. For the other extreme, a 2D crystal symbolized by *E* symmetry could only be coincide with itself after 360° rotation, which means asymmetrical entirely (Fig. 1).

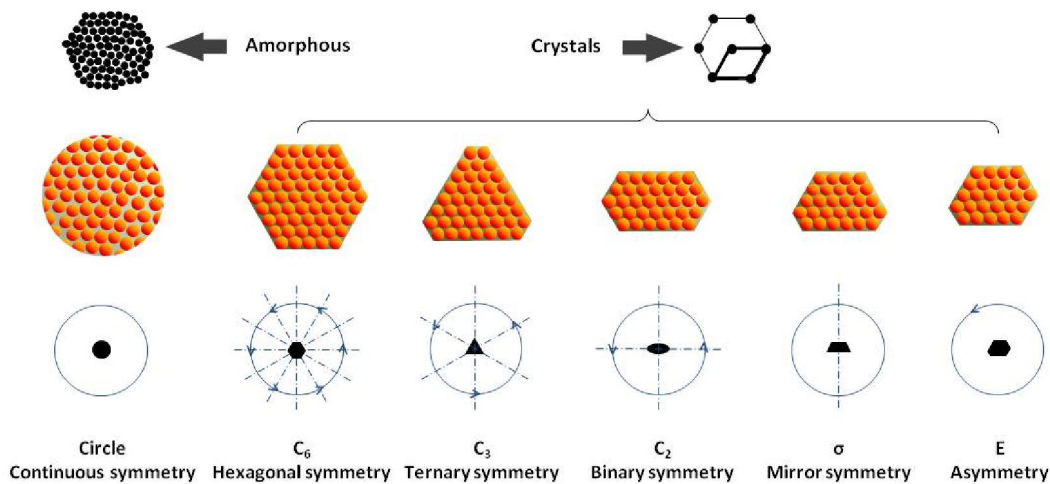

**Figure 1** Systemic symmetry of 2D structures: amorphous wafer with continuous circle symmetry and crystal particles with different systemic symmetry based on the same cell and lattice.

Similarly, a 3D crystal particle possesses systemic symmetries, too. Here we define the 'systemic symmetry' for crystals: when the whole structure of the relatively isolated crystal is performed systemic symmetry operations, such as rotation, reflection or inversion, the orientation and the lattice of the structure are indistinguishable accurately, while the figure of the crystal particle is coincidental approximately with the state in origin. As for the inner defects, they are ignored completely. The duplicity of accuracy and fuzziness characterize the systemic symmetry for crystals.

This definition of systemic symmetry could be interpreted as contrary to the belief that the carriers of symmetries are no more than the cell and lattice in crystallography currently. For that, it is necessary to review an elementary feature of this concept and to show how it naturally leads to the consideration of symmetry breaking, since in general, symmetry tends to decline spontaneously during its development.

## 2. Applications: systemic symmetry and symmetry breaking during nanocrystallization

We reexamine in this light some empirical observations, and propose to interpret them as implying, for two examples, that the aeroplane-like ZnO nanocrystals are caused by spontaneous systemic symmetry breaking and the tetrahexahedral Pt nanocrystals are caused by symmetry breaking during Landau phase transition, though various speculated mechanisms have been proposed or interpreted for aeroplane-like ZnO and tetrahexahedral Pt nanocrystals. Thus, the so far esoteric concept of systemic symmetry breaking is shown to have simple and concrete applications and great usefulness.

## 2.1 Systemic symmetry breaking during the formation of aeroplane-like ZnO nanocrystals

Returning to the first question in the topic, let's use the systemic symmetry and symmetry breaking to attempt investigating the formation of the aeroplane-like ZnO nanocrystals.

The aeroplane-like ZnO nanocrystals is a single crystal with morphology resembling a folded aeroplane[1], and its formation process is interpreted as Fig. 2. First, a tetrapod structure forms and then develops into a tetrapod with 6 sheets between any two legs. Here the tetrapod and the developed tetrapod are both twinning crystals with systemic symmetry of tetrahedron group, for each of them includes four indistinguishable parts, just as that a regular tetrahedron possesses. Second, since the growth of the crystal will involve higher energy for the increasing twinning planes, the twinning crystal may recrystallize into a single one to reduce the energy of the state. Here the problem aroused by recrystallization, the dominating orientation among the four equivalent candidates in the twinning, is thus needed to discriminate. We propose to interpret this situation as implying, for example, that the chlorination from $CH_4$ to $CH_3Cl$ involves four equivalent routes while the result occupies randomly only one of them. For both the twinning and the $CH_4$, the symmetry of tetrahedron group ($T_d$) develops into that of triangle group ($C_3$). Yet all of the developed states, including the other three states not being occupied, can—in a statistic sense—be considered remaining tetrahedron symmetry, for each of these equivalents will take place with equal probability of 0.25. Thus, this symmetry breaking results in an intermediate with such a structure: a rod, its adjacent wire and three uniformly distributed sheets compose a single crystal, while the other three single crystal wires with different orientations are connected to the end of the three sheets via grain boundaries, respectively[2]. Consequently, the growth of the above intermediate develops into an entirely single crystal with aeroplane-like morphology and $C_3$ symmetry.

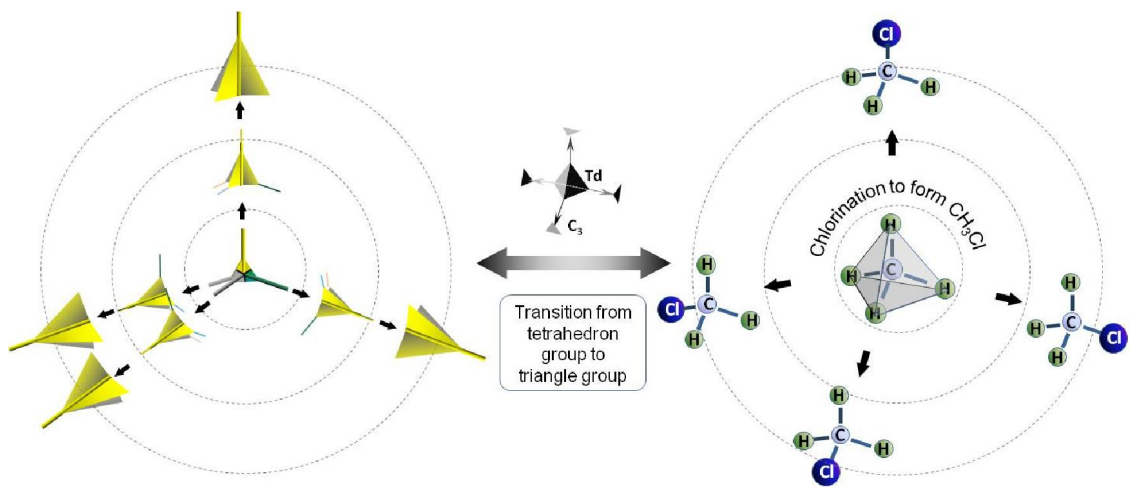

**Figure 2** The transformation from a tetrapod structure to an aeroplane-like structure resembles the chlorination from $CH_4$ to $CH_3Cl$.

In physics, the transition above is defined as spontaneous symmetry breaking that the system transforms spontaneously from a higher symmetrical state to one of the multi-folded states with lower symmetry. In fact, spontaneously symmetry breaking occurs naturally in many situations, and here it is verified empirically by the nanocrystallization of aeroplane-like ZnO particles obtained in our one-pot sintering (Fig. 3a), which follows the systemic symmetry breaking.

Unfortunately, a scientist may be seldom concerned with the systemic symmetry of a nanocrystals (hence the systemic symmetry breaking), and he may suspect the formation of aeroplane-like structure as shown in Fig. 3b. In fact, this mechanism seems to be more acceptable and reasonable for many people even after it has been refuted down[3].

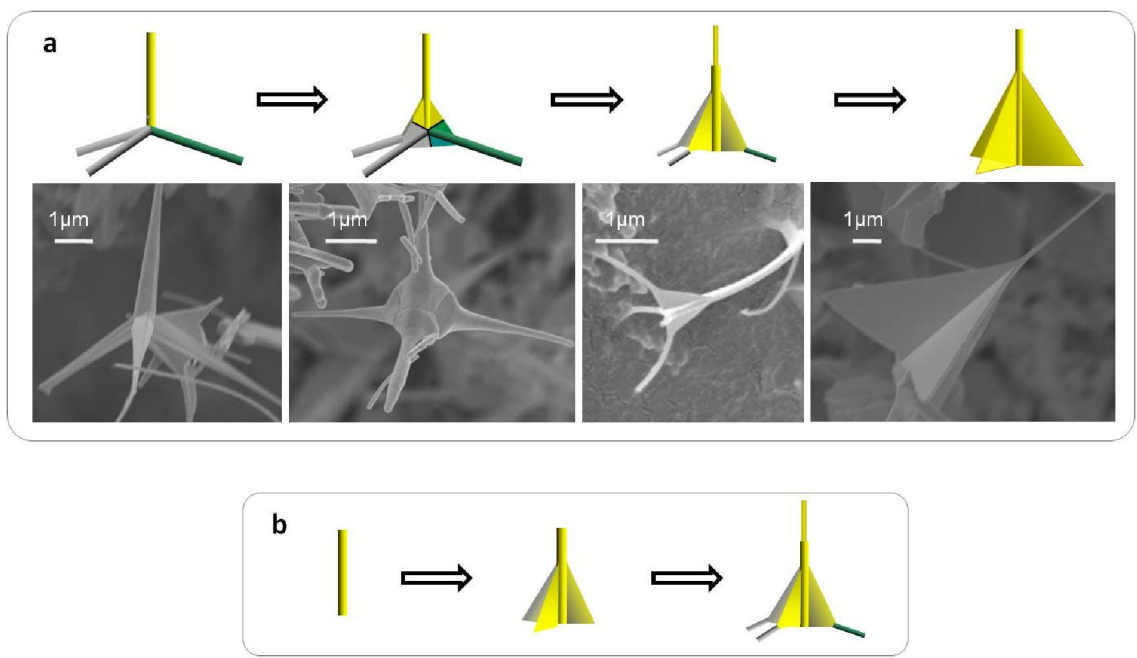

**Figure 3** Results of one-pot experiment prove the mechanism of systemic symmetry breaking (**a**) and refute the *Third Mechanism for ZnO tetrapod nanocrystals* (**b**)[2].

**2.2 Systemic symmetry breaking during the formation of tetrahexahedral Pt nanocrystals**

Returning to the second question in the topic, let's use the systemic symmetry and symmetry breaking to attempt investigating the formation of the tetrahexagonal Pt nanocrystals. Tetrahedral Pt nanocrystals attract more attention for their unusual shape and high catalytic activity[4]. Their surface facets, which were named as {730}, {520} or {210} in different crystals according to different papers, are stable thermally and present constantly in contrast with their various designations[4-7].

We interpret the formation of these unusual facets as the mechanism shown in Fig. 4. The process begins with an amorphous Pt ball under possible conditions[4], and this nanoparticle is continuously spherical symmetry for its isotropy (Fig. 4a). Then the subsequent crystallization will transform the amorphous sphere to a crystal. The systemic symmetry of the Pt crystal, considered together with the symmetry of its cell (fcc) and lattice ($Pm\bar{3}m$), should be as symmetrical as a cube/octahedron ($O_h$ group) for its anisotropy. Deducing directly the formation of crystal ball will involve a problem that the indices of the crystal's surface are almost various continuously and with the unequal surface energy everywhere. Thus, it is necessary for the surface to tune itself to meet two requirements to reach a stable and reasonable state, one is to equalize the surface energy; the other is to obtain the $O_h$ symmetry. These two limits suggest three points: firstly, only the 'inner' atoms of the crystal could be 'solid' on the knots of the lattice, while the other atoms in or near the surface would be 'liquid' till they reach the stable state; secondly, here this suppositionally stable state will be easier to reach when the 'liquid' atoms are divided into some unconnected topologically zones, for every zone could tune itself imperviously to others; lastly, the integration of these divided zones should be $O_h$ symmetrical.

The only route satisfied the above requirements is shown in Fig. 4b-e. Calculation shows, when the length (L) of the suppositional 'solid' reaches the critical value of $\sqrt{2}r$, the surface of the sphere could be divided into 6 parts unconnected topologically (Fig. 4c-d); each of the zone attached to a structure of square symmetry ($C_4$ group) will tune itself till reaching the final state with $C_4$ symmetry (Fig. 4e). Thus, each of the zone tunes itself to a pyramid finally, and the particle ends with a tetrahedral Pt nanocrystal as shown in Fig. 4f.

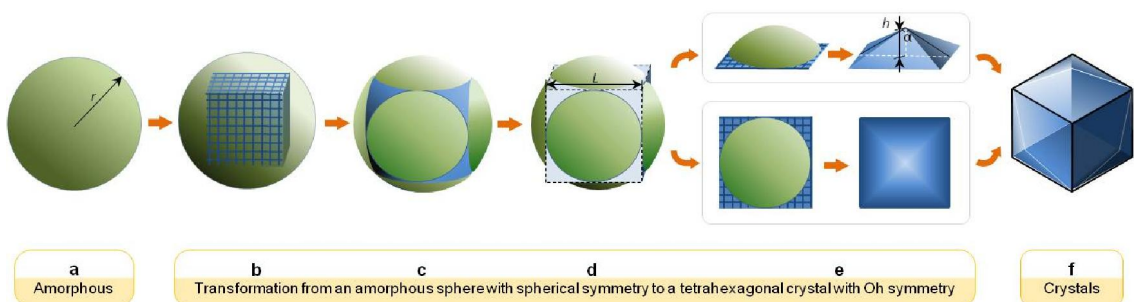

**Figure 4** Scheme of mechanism for the formation of Pt tetrahexahedron.

Consequently, it would be nice to have a calculation linking the above analysis with the empirical results (parameters including $r$, $L$, $h$ and $\alpha$ are shown in Fig. 4): when $L$ reaches $\sqrt{2}r$, $h=(\frac{\pi}{6}-\frac{\sqrt{2}}{2})r$, and $tg(\frac{\alpha}{2})=\frac{L}{2h}=1:(\frac{\sqrt{2}}{3}\pi-1)=2.08= 7 : 3.36 = 5 : 2.40 = 2 : 0.96$. Accordingly, the angles in the perfect octahedron are $\alpha$ =128.7 ° and $\beta$ =141.3 °. Fig. 5 shows the comparison between the real structure [8] and the calculation.

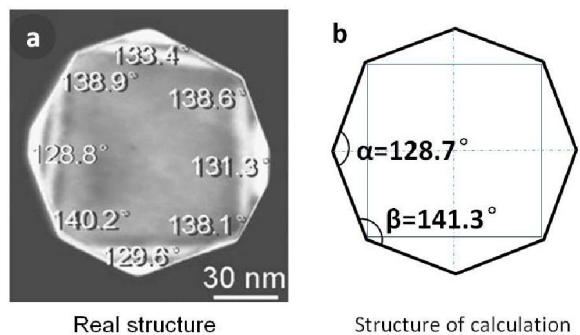

Real structure             Structure of calculation

**Figure 5** Comparison between real structure (**a**) of Pt tetrahexahedron[8] and calculation (**b**).

In physics, Landau theory interpreted well the phase transition between a liquid and a crystal as symmetry breaking[9], which here reduces the continuous symmetry of the amorphous ball to the discrete symmetry of the crystal ($O_h$ group).

Thus, this case involving symmetry breaking during Landau phase transition discovers what leads us to take the facets as {730}, {520} or {210}, and, to speak of the high index of these facets is naturally meaningless, for what we read off from the real structure are just a series of facets based on the asymptotic surface of $\{(\frac{\sqrt{2}}{5}\pi-1),1,0\}$.

### 3. Conclusions

The reexamination of empirical observations, including aeroplane-like ZnO nanoparticles and tetrahexahedral Pt Nanocrystals, indicates that systemic symmetry is a potent tool in the study of nanocrystallization. Practical application of the notion of systemic symmetry requires further considerations, because crystals with perfectly systemic symmetry are seldom encountered in nature. However, crystals with fuzzily systemic symmetry are often encountered at meso-scale. We attribute these phenomena to thus: micro-nano crystals always form within very short time and very small space, and such an environment tends to be an approximately symmetrical and constant one, which introduces litter external cause to break the systemic symmetry. Once the systemic symmetry is recognized, the concept of symmetry breaking, which implies that the group of the systemic symmetry transforms to its subgroup step by step, may be further generalized.

To specify the physical conditions for the systemic symmetry breaking during nanocrystallization is not a fully solved problem. In fact, even the idea that to evaluate the systemic symmetry for a not-so-isolated crystal particle raises a number of conceptual problems, such as the effects of not-so-symmetrical environment, including temperature field, concentration field, gravity field, and so on. Therefore, for the systemic symmetry breaking during nanocrystallization, the most that can be said with perfect safety is that it is compatible with the other laws of crystallographic theories. Empirical scientists having to be content with the products controlled by these laws, we favor the systemic symmetry breaking stated in this letter.

**References**

1. Liu, F., Cao, P., Zhang, H., Li, J. & Gao, H. Controlled self-assembled nanoaeroplanes, nanocombs, and tetrapod-like networks of zinc oxide. *Nanotechnology* **15**, 949 (2004).
2. Gong, J., *et al*. A third kind growth model of tetrapod: Rod-based single crystal ZnO tetrapod nanostructure. *Materials Chemistry and Physics* **112**, 749-752 (2008).
3. Zhou, Y., Zhang, J., Li, J. & Zhang, B. Spontaneous symmetry breaking during formation of ZnO nanocrystals. *Journal of Crystal Growth* (2011).

4. Tian, N., Zhou, Z.Y., Sun, S.G., Ding, Y. & Wang, Z.L. Synthesis of tetrahexahedral platinum nanocrystals with high-index facets and high electro-oxidation activity. *Science* **316**, 732 (2007).
5. Tian, N., Zhou, Z.Y. & Sun, S.G. Platinum metal catalysts of high-index surfaces: from single-crystal planes to electrochemically shape-controlled nanoparticles. *The Journal of Physical Chemistry C* **112**, 19801-19817 (2008).
6. Tian, N., Zhou, Z.Y., Yu, N.F., Wang, L.Y. & Sun, S.G. Direct Electrodeposition of Tetrahexahedral Pd Nanocrystals with High-Index Facets and High Catalytic Activity for Ethanol Electrooxidation. *Journal of the American Chemical Society* **132**, 7580-7581 (2010).
7. Ming, T., *et al.* Growth of Tetrahexahedral Gold Nanocrystals with High-Index Facets. *Journal of the American Chemical Society* **131**, 16350-16351 (2009).
8. Ding, Y., *et al.* Facets and surface relaxation of tetrahexahedral platinum nanocrystals. *Applied Physics Letters* **91**, 121901 (2007).
9. Landau, L. On the theory of phase transitions, part I. *Sov Phys JETP* **7**, 19ff (1937).

## Acknowledgements

The authors greatly acknowledge the NSFC (No. 50283008, 50721062) for financial support.